\documentclass[twocolumn,trackchanges,tighten]{aastex63}

\usepackage{amsmath}
\usepackage{xcolor}

\usepackage{soul}

\usepackage{soul}

\received{}
\revised{}
\accepted{}
\submitjournal{ApJ}

\shorttitle{}
\shortauthors{Krapp, Kratter}

\begin{document}

\title{A thermodynamic criterion for the formation of Circumplanetary Disks}

\correspondingauthor{}
\email{krapp@arizona.edu}

\author[0000-0001-7671-9992]{Leonardo Krapp}
\affiliation{Department of Astronomy and Steward Observatory,  University of Arizona, Tucson, Arizona 85721, USA}
\affiliation{Departamento de Astronomía, Facultad de Ciencias Físicas y Matemáticas Universidad de Concepción, Av. Esteban Iturra s/n Barrio Universitario, Casilla 160-C, Chile}

\author[0000-0001-5253-1338]{Kaitlin M. Kratter}
\affiliation{Department of Astronomy and Steward Observatory,  University of Arizona, Tucson, Arizona 85721, USA}

\author[0000-0002-3644-8726]{Andrew N. Youdin}
\affiliation{Department of Astronomy and Steward Observatory, University of Arizona, Tucson, AZ 85721, USA}
\affiliation{The Lunar and Planetary Laboratory, University of Arizona, Tucson, AZ 85721, USA}

\author[0000-0002-3728-3329]{Pablo Ben\'itez-Llambay}
\affiliation{Facultad de Ingenier\'ia y Ciencias, Universidad Adolfo Ib\'añez, Av. Diagonal las Torres 2640, Peñalol\'en, Chile}

\author[0000-0002-9626-2210]{Fr\'ed\'eric Masset}\affiliation{Instituto de Ciencias F\'isicas, Universidad Nacional Aut\'onoma de M\'exico, Av. Universidad s/n, 62210 Cuernavaca, Mor., Mexico}

\author[0000-0001-5032-1396]{Philip J. Armitage}
\affiliation{Center for Computational Astrophysics, Flatiron Institute, 162 Fifth Avenue, New York, NY 10010, USA}
\affiliation{Department of Physics and Astronomy, Stony Brook University, Stony Brook, NY 11794, USA}

\begin{abstract}
The formation of circumplanetary disks is central to our understanding of giant planet formation, influencing their growth rate during the post-runaway phase and observability while embedded in protoplanetary disks. We use 3D global multifluid radiation hydrodynamics simulations with the FARGO3D code to define the thermodynamic conditions that enable circumplanetary disk formation around Jovian planets on wide orbits. 
Our simulations include stellar irradiation, viscous heating, static mesh refinement, and active calculation of  opacity based on multifluid dust dynamics. 
We find a necessary condition for the formation of circumplanetary disks in terms of a mean cooling time: 
when the cooling time is at least one order of magnitude shorter than the orbital time scale, the specific angular momentum of the gas is nearly Keplerian at scales of $R_{\rm{Hill}}/3$.
We show that the inclusion of multifluid dust dynamics favors rotational support because dust settling produces an anisotropic opacity distribution that favors rapid cooling. 
In all our models with radiation hydrodynamics, specific angular momentum decreases as time evolves in agreement with the formation of an inner isentropic envelope due to compressional heating.
The isentropic envelope can extend up to $R_{\rm{Hill}}/3$ and shows negligible rotational support.
Thus, our results imply that young gas giant planets may host spherical isentropic envelopes, rather than circumplanetary disks.
\end{abstract}


\section{Introduction} \label{sec:intro}
In the dominant planet formation paradigm of core accretion \citep{Pollack1996}, gas giants grow through accretion of solids and gas from the protoplanetary disk. At early times mass accretion onto the envelope is limited by Kelvin-Helmholtz cooling, which radiates away energy from gravitational contraction and accretion of solids; as the planet cools, more mass can become bound. In this phase, the outer boundary of the planetary atmosphere is matched to either the Hill Radius or Bondi Radius, whichever is smaller \citep[e.g.,][]{Pollack1996, Ayliffe2009I, Tanigawa2012,Ormel2015,Cimerman2017, Kurokawa2018,Lambrechts2017a,Bethune2019,Zhu2021}. A sharp transition in the evolution to the so-called {``runaway" phase} occurs when the envelope becomes strongly self-gravitating \citep{Pollack1996, Hubickyj2005,Rafikov2006,Dangelo2008,Piso2014,Lee2015,DANGELO2021}. The envelope contraction rate accelerates, allowing mass to accrete onto the envelope at an increasing rate. 
This phase of growth is expected to continue until limited by the supply of gas from the disk.

Subsequent to the runaway phase, the envelope is thought to accrete at the rate supplied by the disk, which depends upon disk processes including gap opening and photoevaporation that reduce the mass supply \citep{Lissauer2009, Mordasini2012,DAngelo2013}. The accretion of angular momentum from the disk, in excess of what can be accommodated within the spin angular momentum of the planet, leads to the expectation that circumplanetary disk formation will also happen at this time \citep{Quillen_1998,Ward2010}. This process has been studied with a range of methodologies from 1D solutions of the stellar-structure equations to fully three-dimensional radiation hydrodynamic simulations  \citep{Klahr2006,Marley2007,Spiegel2012,Mordasini2012,Berardo2017,Aoyama2021,Lambrechts2019,Fung2019,Szulagyi2016,Szulagyi2019,Schulik2019,Marleau2023,Choksi2023}.

Existing three-dimensional simulations have shown that post-runaway accretion can occur via either a distinguishable circumplanetary disk or via a weakly rotationally supported envelope, depending upon the thermal state of the circumplanetary gas. In the simplest (and best understood) limits, isothermal models produce robust rotationally supported disks, while adiabatic systems host weakly rotationally supported envelopes \citep{Fung2019}. More physically, this bifurcation in flow morphology can be related to the opacity, with pioneering work by \citet{Ayliffe2009II} concluding that sufficiently low opacity enables strong rotational support. Follow-up work by \cite{Szulagyi2016} and \cite {Szulagyi2017} instead frames the circumplanetary disk formation question in terms of temperature at the inner boundary of the simulation and planet mass. 
 A correlation between opacity, planet mass, and average azimuthal velocity has also been discussed in \cite{Schulik2020}.
In this work, we demonstrate a general correlation between specific angular momentum and cooling time, which is agnostic to the planetary and disk parameters. Thus we circumvent framing the question of circumplanetary disk formation in terms of planet mass, disk opacity, and disk initial conditions. 

For black hole accretion, the dividing line between the geometrically thin disks found in Active Galactic Nuclei, and the thick flows inferred for the Galactic Center and M87, is understood to depend upon how fast the gas can lose energy as it accretes \citep{Rees82}. 
In this work, we seek an analogous thermodynamic criterion for circumplanetary disk formation in terms of the average cooling time and the specific angular momentum. Because of the primacy of opacity in setting the cooling as a function of location and the low temperatures of the disk, it is crucial to include dust dynamics \citep{Isella2018,Chachan2021,Krapp2022}. We develop and validate our cooling criterion using global, radiation hydrodynamic simulations of Jovian-mass planets in the outer regions of protoplanetary disks. The simulations include full multi-species dust transport that determines an actively evolving opacity, allowing us to account for how the variations in opacity due to the settling of different-size particles modify the cooling time. We show how consideration of the cooling rate as a function of the radius within the Hill sphere both encompasses prior results and provides a more general criterion for circumplanetary formation that incorporates temperature, density, opacity, and planet mass.

\medskip
This work is organized as follows. In Section\,\ref{sec:numerical} we introduce our numerical framework. In Section\,\ref{sec:Results} we describe the main results of our numerical exploration with emphasis on the evolution of cooling time and specific angular momentum inside the Hill sphere. In Section\,\ref{sec:discussion} we discuss our findings and summarize our work.

\begin{figure*}[t]
    \centering
    \includegraphics[]{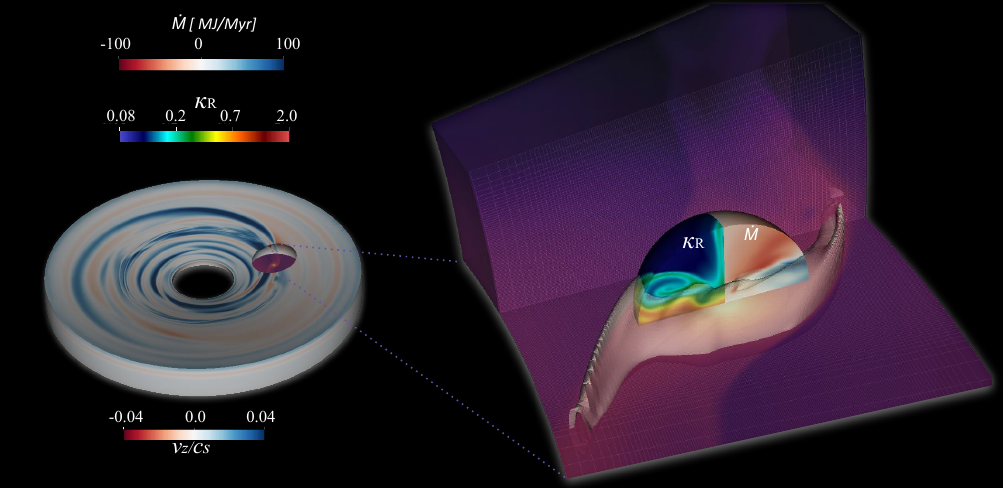}
    \caption{Global \texttt{multifluid} model after 2 orbits. The local patch shows the opacity and mass flux (in units of Jupiter mass per year) at the scales of the Hill sphere. The background magenta color displays the gas density for reference. 
    The inclusion of dust dynamics introduces an anisotropic opacity distribution in the radiative transfer model, which agrees with the opacity distribution estimated with isothermal models in \cite{Krapp2022}. 
    The mass flux balances between polar inflow (red color) and midplane outflow (blue color). 
    The surface contour corresponds to an isodensity with $\rho_{\rm g}=3.5 \times 10^{-12}\,{\rm g}{\rm cm}^{-3}$ that maps the initial envelope. This envelope will become more spherically symmetric as time evolves and rotational support is weakened. The global disk on the left shows the vertical velocity of the gas. The static mesh refinement allows us to solve the dynamics inside the Hill sphere as well as the large-scale flow exited by the planet.   }
    \label{fig:fig1}
\end{figure*}
\section{Numerical Method}
\label{sec:numerical}
The numerical simulations in this work are carried out using the multi-fluid code FARGO3D \citep{Benitez-Llambay2016,Benitez-Llambay2019}. 
We first introduce the equations in Section\,\ref{sec:equations}, then continue with a summary of the key features of the numerical method in Section \ref{sec-fargonum}. 

\subsection{Equations}
\label{sec:equations}

Our hydrodynamics simulations solve the following set of equations for the gas and dust fluid.
\begin{eqnarray}
 \partial_t \rho_{\rm g} + \nabla \cdot\left( \rho_{\rm g} \mathbf{v}_{\rm g}\right) & = & 0\,, \label{eq:gas_cont}  \\
  \partial_t \rho_{{\rm d}j} + \nabla \cdot\left( \rho_{{\rm d}j} \mathbf{v}_{{\rm d}j} + {\bf j}_{{\rm d}j}\right) & = & 0\,, \\
  \partial_t \mathbf{v}_{\rm g} + \mathbf{v}_{\rm g} \cdot \nabla \mathbf{v}_{\rm g} & =& - \frac{\nabla P}{\rho_{\rm g}} - \nabla \Psi +\frac{1}{\rho_{\rm g}}\nabla \cdot {\bf \tau} \nonumber \\
  &-& \frac{\Omega}{\rho_{\rm g}} \sum^N_{j=1} \frac{\rho_{{\rm d}j}}{T_{{\rm s}j}} \left( {\bf v}_{\rm g} - {\bf v}_{{\rm d}j} \right), \\
   \partial_t \mathbf{v}_{j} + \mathbf{v}_{j} \cdot \nabla \mathbf{v}_{j}  & = & - \nabla \Psi  - \frac{\Omega}{T_{{\rm s}j}}\left( {{\bf v}_{{\rm d}j}-\bf v}_{\rm g} \right), \label{eq:dust_mom}
\end{eqnarray}
for $j=1\dotsc N-1$. $\rho_{\rm g}$, $\rho_{{\rm d}j}$, ${\bf v}_{\rm g}$ and ${\bf v}_{{\rm d}j}$ correspond to the gas and dust densities and velocities, respectively.
The angular velocity $\Omega = \sqrt{GM_\odot/R^3}$.
The dust diffusion flux is given by
\begin{equation}
\label{eq:diffusion}
{\bf j}_{{\rm d}j} = -D \left( \rho_{\rm g} + \rho_{{\rm d}j}\right) \nabla \left(\frac{\rho_{{\rm d}j}}{\rho_{\rm g} + \rho_{{\rm d}j}}\right)\,,
\end{equation}
where $D$ is the diffusion coefficient. The Stokes number corresponds to
\begin{equation}
\label{eq:size}
T_{{\rm s}j}  = {\rm{min}}\left( a_j \sqrt{\pi/8} \frac{\rho_{\rm solid} \Omega}{\rho_{\rm g} c_{\rm s}}, 1.5 \right)\,,
\end{equation}
with $a_j$ and $\rho_{\rm solid}=1.6686\,{\rm g}/{\rm cm}^3$ the particle size and material density, respectively. In all our disk models, the maximum Stokes number at the vicinity of the planet is $T_{{\rm s}} \lesssim 1$. 
To prevent large Stokes numbers at very low-density regions we force the maximum Stokes number to $T_{{\rm s}}=1.5$.  

The gravitational potential $\Psi$ includes the contributions from the central star and the planet and neglects the indirect term, thus
\begin{equation}
    \Psi = -\frac{GM_\odot}{R} - \frac{GM_{\rm p}}{\sqrt{|{\bf R} - {\bf R}_{\rm p}|^2 + R^2_s}}\,,
\end{equation}
where  $R_s$ is a softening length, and $M_{\rm p}$ and ${\bf R}_{\rm p}$ correspond to planet mass and radial vector position. In this work, we consider a Jupiter mass-like planet on a fixed circular orbit at $R_p = 30 {\rm AU}$ from the central star, with orbital frequency $\Omega_p = \sqrt{GM_\odot/R^3_p}$.
The viscous stress tensor, ${\bf \tau}$, is given by
\begin{equation}
    {\bf \tau} = \rho_{\rm g} \nu \left( \nabla {\bf v} + \left(\nabla {\bf v}\right)^{T} - \frac{2}{3} (\nabla \cdot {\bf v}) {\bf 1} \right)\,,
\end{equation}
with $\nu$ the gas kinematic viscosity. 

The gas energy is obtained from radiative transfer equation based on the Flux-Limited-Diffusion approximation \citep{Krumholz2007},
\begin{eqnarray}
  \partial_t {e}_{\rm g} + \nabla \cdot( {e}_{\rm g} \mathbf{v}_{\rm g} ) & =& - P \nabla \cdot {\bf v}_{\rm g} - \rho \kappa_{\rm p}(4\sigma T^4 - c E_{\rm R}) + Q^s + Q^+\,,  \nonumber  \\
 \partial_t  {E}_{\rm R} + \nabla \cdot {\mathbf F}_{\rm R}& =&  \rho \kappa_{\rm p}(4\sigma T^4 - c E_{\rm R})\,, 
  \label{eq:energy}
\end{eqnarray}
with $e_{\rm g} = c_v \rho_{\rm g} T$ the gas internal energy, $T$ the gas temperature, $\kappa_{\rm P}$ the Planck mean opacity, $\kappa_{\rm R}$ the Rosseland mean opacity, $\sigma$ the Stefan-Bolztmann constant, $E_{\rm R}$ the radiative energy, and ${\bf F}_{\rm R}$ the radiative flux which is obtained as
\begin{equation}
    \mathbf{F}_{\rm R} = -\lambda \frac{c}{\rho \kappa_{\rm R}} \nabla E_{\rm R}\,,
\end{equation}
where  $c$ is the speed of light, and $\lambda$ is the flux-limiter defined as \citep{Levermore1981}
\begin{equation}
\lambda = 
     \begin{cases}
       \frac{2}{3+\sqrt{9+10 \mathcal{P}^2}} &\quad\text{if\,} \mathcal{P} \leq 2\\
      \frac{10}{10 \mathcal{P} + 9 + \sqrt{81 + 180 \mathcal{P}}} &\quad\text{if\,} \mathcal{P}>2\\ 
     \end{cases}
\end{equation}
with $\mathcal{P} = |\nabla E_{R}|/(\rho_{\rm g} \kappa_{\rm R} E_{\rm R})$.
In this work, we assume $\kappa_{\rm R}=\kappa_{\rm P}$ since they show little discrepancies at the scale of the Hill sphere in the calculations of \cite{Krapp2022}. The Planck mean opacity is obtained utilizing the dust densities as
\begin{equation}
    \label{eq:extinction}
    \kappa_{\rm P} = \frac{\sum^{j=N-1}_{j=1} \rho_{\rm d}(a_j)\,\kappa(a_j)  }{ \rho_{\rm g}+\sum^{j=N-1}_{j=1} \rho_{\rm d}(a_j) }\,.
\end{equation}
where $\kappa(a_j)$ is the obtained from the approximate emissivity averaged
over the Planck spectrum \citep{Barranco2018} 
\begin{equation}
\kappa(a_j) = \frac{{\rm min}\left(1, \frac{T a_j}{600 \mu{\rm m K}}\right)}{4/3 a_j \rho_{\rm solid}}\,.
\end{equation}
stellar, $Q^s$, and viscous, $Q^{+}$, heating terms in Eq.\,\eqref{eq:energy} are implemented following \cite{Bitsch2013}.

\subsection{ $\rm{FARGO3D}$ numerical simulations} \label{sec-fargonum}

We solve Eqs.\,\eqref{eq:gas_cont}-\eqref{eq:dust_mom} using the code FARGO3D with nonuniform meshes and the rapid advection algorithm for arbitrary meshes (RAM) \citep{BenitezLlambay2023}. 
We perform global three-dimensional simulations on a spherical mesh centered at the star, with coordinates $(R,\varPhi,\Theta)$.  
The numerical domain comprises only half of the disk in the vertical direction, with $\Theta \in[0.44\pi,\pi/2]$. The radial domain is set to $R\in[0.4,2.0]$ whereas in azimuth we include the entire disk $\varPhi \in [-\pi,\pi]$. 
Results will be described as a function of spherical coordinates centered at the planet defined as $(r, \phi, \theta)$. 

We use a nonuniform static mesh to increase the numerical resolution in the vicinity of the planet. 
The grid is obtained using a mesh density function method. For azimuth (and latitude) and radial direction, the mesh density functions follow from equations B1 and B8 of \cite{BenitezLlambay2023}, respectively.  The parameters of the mesh density functions are shown in Table\,\ref{table:mesh}.

The planet potential is fixed at $R=1.0$, $\varPhi=0$ and $\Theta=\pi/2$. In this work, we consider a planet mass of $M_p = 10^{-3} M_\odot$.
The Hill radius corresponds to $r_H = R_p\left( M_p/(3M_{\odot})\right)^{1/3}$. 
The softening length defines a sphere with a radius of $\sim 2$ grid cells centered at the planet, which gives $R_s = 0.0039 R_p$ (about $\%5$ of the Hill radius) for our standard resolution.
\begin{table}[]
\begin{center}
		\caption{Parameters of the Mesh density functions.
		\label{table:mesh}}
\begin{tabular}{lcccccc}
	\decimals
	\hline
	\hline
Coord.  & Center & Min/Max & a & b & c & $N_{\rm cells}$  \\
\hline
$\varPhi$    &  0     &  $-\pi/\pi$ & $0.5r_{\rm H}$  & $1.5r_{\rm H}$& 15 & 512 \\
$\Theta$  &  $\pi/2$  &  $(1.38)/(\pi/2)$ & $0.5r_{\rm H}$  & $2.0r_{\rm H}$& 8 & 84 \\
$R$  &  $1.0$  &  $0.4/2.0$ & $0.5r_{\rm H}$  & $2.0r_{\rm H}$& 9 & 260 \\
\hline
  \end{tabular}
\end{center}
\tablecomments{\footnotesize The parameters $a$, $b$ and $c$ define the refinement level utilizing the mesh density functions from \cite{BenitezLlambay2023}, whereas $r_H$ is the Hill radius of a Jupiter-mass-like planet at 30 AU from the central star.} 
\end{table}

\subsubsection{Boundary conditions}

Since we simulate only one disk hemisphere, we adopt reflecting boundary conditions at the disk equator.
At $\Theta=0.44\pi$ we set $\partial_{\Theta} v_{\Theta} =0$ if $v_\Theta < 0$, otherwise $v_\Theta =0$.
For all other variables, we set $\partial_{\Theta}=0$.
We also set the radiative energy to $E_{\rm rad} = 4\sigma T^4_b/c$ with $T_b = 8 {\rm K}$ at $\Theta=0.44\pi$.
In the radial direction, we set the radial derivatives of the density and dust azimuthal velocity are set to zero. 
We additionally include buffer zones that restore the density to be equal to the initial value, which also prevents undesired reflections that may perturb the flow at the region of interest \citep[][]{DeVal-Borro2006}. 
In the inner buffer zone, we set the Rosseland mean opacity to $\kappa_{\rm R}=0.1\, {\rm cm}^2 {\rm g}^{-1}$ to prevent the formation of a density bump due to the strong heating from stellar irradiation.
Finally, we adopt periodic boundary conditions in the azimuthal direction.

\subsubsection{Initial conditions and disk model}
Our initial disk model considers an axisymmetric vertically isothermal protoplanetary disk with gas in hydrostatic equilibrium and dust in settling-diffusion equilibrium as described in \citep{Krapp2022}. 
The initial gas surface density $\Sigma_0(R) =  3.5\left( R/R_p \right)^{-1} \, {\rm g}/{\rm cm}^2 $ and temperature $T(R) = 30 \left( R/R_p \right)^{-1/2}{\rm K}$.
The initial dust surface density corresponds to $\Sigma_{{\rm d}j}(R)=\epsilon_j\Sigma_0(R)$, where the dust-to-gas density ratio is obtained assuming a power-law distribution
\begin{equation}
    \epsilon_j = \epsilon \, \frac{ a^{4-s}_{j+1} - a^{4-s}_{j}}{a^{4-s}_{\rm max} - a^{4-s}_{\rm min} }\,.
\end{equation}
The slope of the distribution corresponds to $s=3.5$ and the total dust-to-gas mass ratio is set to $\epsilon=0.017$. 
The sizes are obtained considering a log-uniform distribution with  $a_{\rm max}=0.4\,{\rm cm}$ and $a_{\rm min}=5\times10^{-5}\,{\rm cm}$. 
We consider 7 dust fluids and each species has a size $a_{j+1}$ (that is the maximum size of each mass bin). The two fluids with smaller sizes are assumed to be well-mixed with the gas, thus, we only integrate the equations of 5 dust fluids in our simulations. 
The gas kinematic viscosity and dust diffusion are $\nu = 2 \times 10^{-6}\, R^{2}_p \Omega_p$ and $D = 1\times10^{-5}\, R^{2}_p \Omega_p$, respectively. 
While the choice of these parameters is somewhat arbitrary without a self-consistent turbulent model, we stress that dust diffusion alters the Rosseland mean opacity \citep{Krapp2022}, and thus impacts cooling. On the other hand, viscosity plays a minor role in our work since the dominant source of disk heating is stellar irradiation at 30au, except for very high viscosities (equivlanet to $\alpha \gtrsim 0.1$). 
The central star is a solar mass star with an effective temperature of $T_{*} = 5500 {\rm K}$ and a radius of $R_*=1.4 \times 10^{11} {\rm cm}$. Before introducing the planet, we run the disk model with stellar irradiation to find the axisymmetric equilibrium for all the quantities.

The balance between stellar irradiation and radiative cooling generates a hotter surface layer and a vertically isothermal disk region, consistent with the outcome of passively irradiated disk models \citep{Chiang1997}. 
The Hill sphere of the planet will be well embedded in this vertically isothermal region.
We take the outcome from the radiative axisymmetric model after 10 orbits to generate the 3D disk initial conditions by expanding density, velocity, temperature, radiative energy, internal energy, and opacity along the azimuthal coordinate. 
In our models, the temperature, density, and velocity field converge to a numerical equilibrium after 10 orbits. 

\begin{figure}
    \centering
    \includegraphics[]{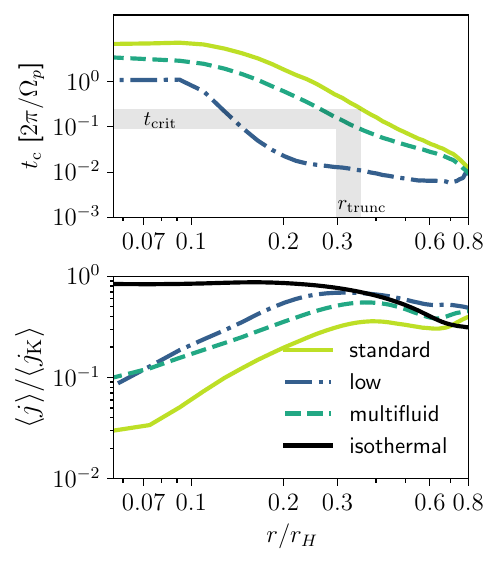}
    \caption{Comparison of effective cooling time and specific angular momentum for the \texttt{standard}, \texttt{multifluid} and \texttt{low} models averaged between 16 and 20 orbits. Gas envelopes with $t_c \gtrsim t_{\rm crit}$ show weak rotational support. This timescale is highlighted with the horizontal dashed line.  We also include a vertical dashed line indicating the tidal truncation radius.  }
    \label{fig:cooling_compare}
\end{figure}

\subsection{Simulation models}

In this work, we consider five different simulations. The models denoted as \texttt{standard}, \texttt{isothermal}, and \texttt{multifluid} correspond to our baseline simulations. The \texttt{standard} model has no dust dynamics whereas the \texttt{multifluid} evolves the opacity self-consistently from dust dynamics. The \texttt{isothermal} model does not include radiative transfer and assumes a local isothermal equation of state with $P=c^2_{\rm s} \rho$. We additionally include a run, \texttt{high\_res} that doubles the resolution of the \texttt{standard} simulation inside the Hill sphere ($110\, {\rm cells}/r_H$). Finally, we will compare the results of the \texttt{standard} model with a reduced opacity model, \texttt{low}, where the dust-to-gas ratio is set to $\epsilon = 3.5\times10^{-4}$.

\subsubsection{Analysis}

To present our results we interpolate the data onto a spherical mesh of radius $r_H$ centered at the planet location. The mesh is evenly spaced with $64^3$ cells. We define the spherical average of a given scalar $f$ as $\langle f \rangle$. We calculate $\langle f \rangle$ on the semi-sphere with radius $r=r_k$ as 
\begin{equation}
    \langle f(r_k) \rangle = \frac{1}{2\pi}\sum_j \sum_i f_{ijk} \sin(\theta_j) \Delta \theta_j \Delta \phi_i\,.
\end{equation}
The total average inside a semi-sphere of radius $r$ is obtained as
\begin{equation}
    \langle \langle f \rangle \rangle_r = \frac{1}{N_r} \sum_k \langle f \rangle\,,
\end{equation}
where $N_r$ denoted the number of grid cells utilized in the interpolation along the radial direction (e.g., $N_r=64$ for $r=r_H$).

\begin{figure*}
    \centering
    \includegraphics[scale=0.9]{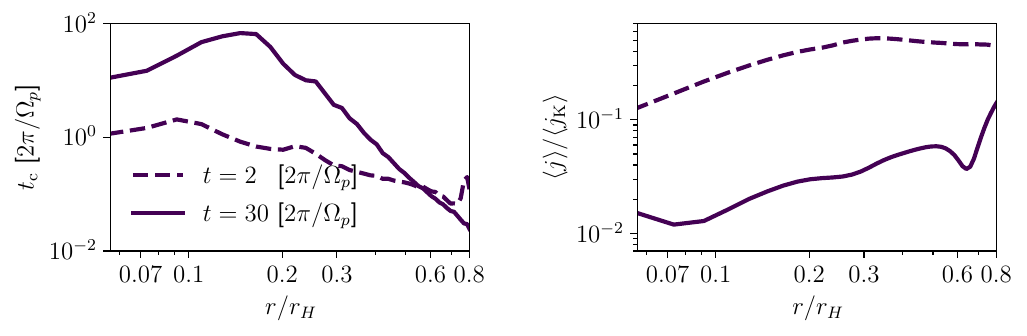}
    \includegraphics[scale=0.6]{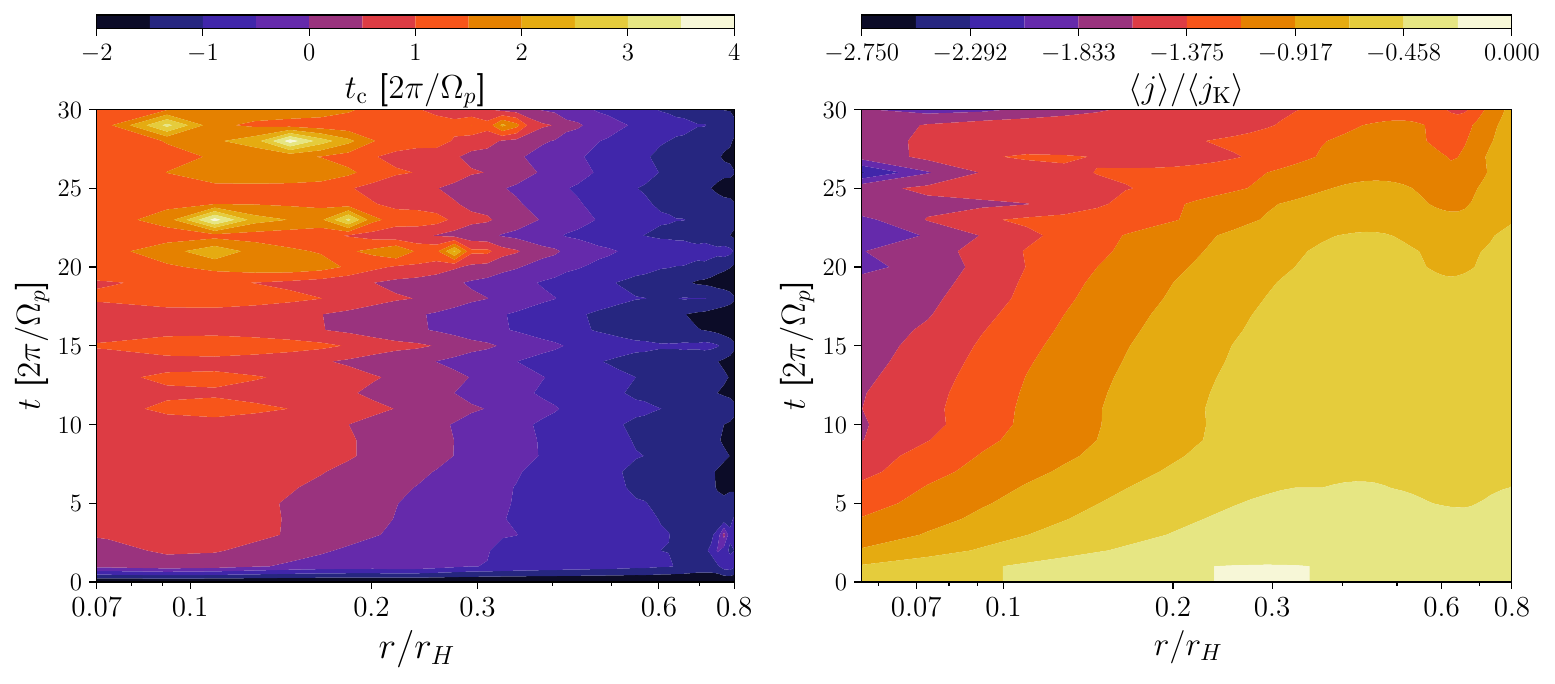} 
    \caption{Top row: Left panel shows the cooling time obtained from Eq.\,\eqref{eq:tcool} as a function of radius inside the Hill sphere. The right panel shows the specific angular momentum normalized by the Keplerian specific angular momentum. Dashed and solid lines correspond to different orbital times. While the cooling time increases inside $r\lesssim
    0.5r_H$, the specific angular momentum decreases, turning the gas into a weakly rotationally supported envelope. Bottom row: cooling time and specific angular momentum as a function of time inside the Hill radius. The cooling time increases as time evolves at all radii inside the Hill sphere. The rather sharp jumps in the cooling time after 5 and 20 orbits are followed by a decay in the specific angular momentum. Results correspond to the \texttt{standard} model. }
    \label{fig:rotational_support}
\end{figure*}
\section{Results}
\label{sec:Results}
In this section, we identify an empirical criterion for the formation of rotationally supported circumplanetary structures. We focus on the correlation between effective cooling time scale and specific angular momentum, which ultimately governs CPD formation. We also provide comprehensive descriptions of the envelope structure and flow patterns observed in our simulation suite. We provide evidence that the slowly evolving three-dimensional flows within the Hill sphere arise due to a subtle balance between accretion and advection of momentum.

In particular, we show that a critical cooling time of only $t_c \gtrsim 0.1\,(2\pi/\Omega_p)$  at scales of $r<r_H/3$ produces weakly rotationally supported envelopes. The choice of $r=r_H/3$ is motivated by the truncation radius in isothermal models of circumplanetary disks \citep{Canup2002,Martin2011}.
That this critical value falls far below unity in the local dynamical timescale illustrates the degree to which even very rapid cooling is not well captured by isothermal runs. We shall show that this criterion is also consistent with previous numerical results from a diverse range of setups.

We also find that rotational support and cooling time in non-isothermal runs are not steady-state quantities that can be assigned to a given planet parameter set. Instead, we find that both quantities continue to evolve over tens to hundreds of orbits. Thus we frame the formation of CPDs not as a dichotomy, but as a time-dependent question.

\subsection{Relating cooling, contraction, and rotational support}

Our simulation suite reveals a correlation between the evolution of the specific angular momentum contained within planetary envelopes, and the cooling rate of the envelope. The balance between the two sets the degree of rotational support in the gas. As the cooling time grows, the luminosity from the contracting atmosphere is increasingly trapped within the envelope, increasing local pressure support. This pressure support helps to drive circulation patterns, which in turn lead to an outward flux of specific angular momentum. Because the quantities evolve slowly in time, our choice of a CPD criterion is based on the measured envelope cooling rate, rather than based on innate parameters of the run.

We estimate the effective radiative cooling time scale of the envelope as 
\begin{equation}
   t_{c} = \frac{ \int^{r_H}_{r} \int^{2\pi}_{0} \int^{\pi/2}_{0}  c_v \rho T r^2{\rm d} r {\rm d} {\phi} \sin(\theta){\rm d}\theta}{  L(r_H) - L(r)  }\,,
   \label{eq:tcool}
\end{equation}
where $c_v$ is the specific heat capacity at constant volume and $L$ is the luminosity of the semi-sphere of radius $r$ and is obtained from the radiative flux as
%
%
\begin{equation}
   L  = -\int^{2\pi}_{0} \int^{\pi/2}_{0} \frac{\lambda c}{\kappa_{\rm R} \rho} \left( \nabla E_{\rm rad} \cdot \hat{r} \right) r^2 {\rm d} {\phi}\sin(\theta){\rm d}\theta\,.
   \label{eq:L}
\end{equation}
Note that the choice of integration limits means that $t_c$ defined here is not the Kelvin-Helmholtz timescale of the contracting protoplanetary envelope in a volume with radius $r$. 
Instead, this definition of $t_c$ captures the cooling timescale of the spherical shell of gas between $r_H$ and $r$.
Thus, $t_c$ indicates how fast material between $r_H$ and $r$ has to cool down to become rotationally supported. With this definition, the cooling time tends to increase towards smaller $r$, which captures the fact that CPDs form from the outside in when the gas is not isothermal.
While formally our Eq.\,\eqref{eq:tcool} includes the Hill surface, in our numerical calculations we will only integrate a volume up to $r=0.9r_H$. 
This choice is only a matter of convenience to avoid the sharp gradients in the luminosity introduced by the shocks of the spiral wakes generated by the planet. 
 
Our cooling time only accounts for the radiative flux, neglecting any contributions from convection (advection of energy through the surfaces). 
In principle one could modify the definition to account for convective energy transport, however in this case one would expect rotational support to be negligible. 

In addition to the effective cooling time we calculate the specific angular momentum\footnote{Note that $j$ corresponds to the vertical component of the specific angular momentum vector along a $z$-axis centered at $R_p$.} of the envelope defined as
\begin{equation}
    j = v_\phi r \sin(\theta)\\.
    \label{eq:jspec}
\end{equation}
 We normalize the specific angular momentum to the Keplerian\footnote{We define the Keplerian specific angular momentum relative to the smoothed potential.} specific angular momentum defined as $j_K = r^2 \sqrt{GM_p} (r^2+R^2_s)^{-3/4}\sin(\theta)$, with 
 \begin{equation}
 \langle j_K \rangle = \frac{\pi \sqrt{GM_p}}{4} r^2 (r^2+R^2_s)^{-3/4}\,.    
 \end{equation}

In Figure\,\ref{fig:cooling_compare} we show the obtained values of $t_c$ and $\langle j \rangle$ averaged between 16 and 20 orbits  for the \texttt{multifluid}, \texttt{low} and \texttt{standard} models.
The values displayed in Figure\,\ref{fig:cooling_compare} are not in steady-state. In our radiation hydro simulations, the specific angular momentum decreases as time evolves while cooling time increases. This is not the case in the isothermal run, where steady-state is reached after 5-10 orbits. Moreover, 
the \texttt{isothermal} model provides a useful upper bound on the specific angular momentum $j$ contained within the envelope (for fixed planet parameters),  since pressure support is minimal. 
In this case, we find that  $\langle j_{\rm iso }\rangle = 0.7 \langle j_K \rangle$ at $r \sim r_H/3$.  This degree of support corresponds to a canonical thick, circumplanetary disk-like morphology in the gas.
Thus in assessing the dividing line between relatively strong and weak rotational support, we look for the values of $t_c$ such that $\langle j \rangle \sim  \langle j_{\rm iso}\rangle$. 
Overall, we find that
\begin{equation}
    \label{eq:tcrit}
    \frac{t_c}{2\pi/\Omega_p} \gtrsim 0.1 \implies \langle j \rangle \lesssim  \langle j_{\rm iso}\rangle\,,
\end{equation}
Eq.\,\eqref{eq:tcrit} sets a thermodynamic criterion for the presence of substantially rotational-supported envelopes at scales of $r\simeq 0.3r_H$. 
In fact, our results suggest that $\langle j \rangle \sim  \langle j_{\rm iso}\rangle$ when $t_c \sim 0.01\, \left(2\pi/\Omega_p\right)$. 

In Figure\,\ref{fig:cooling_compare} we show that at scales of $r \sim r_H/3$, the specific angular momentum $j \simeq 0.75 \langle j_{\rm iso} \rangle$ when $t_c \lesssim 0.1\, [2\pi/\Omega_p]$ for \texttt{multifluid} and \texttt{low} after 20 orbits. However, the \texttt{standard} model shows lower specific angular momentum with a slightly longer cooling time. While the isothermal model has reached a steady state, the specific angular momentum continuously decreases as cooling time increases in the rest of the models. This lack of steady-state can be seen in Figure\,\ref{fig:rotational_support}, where in the top row we show the obtained values of $t_c$ and $\langle j \rangle$ after 2 and 30 planetary orbits for the \texttt{standard} model. 
To further validate the criterion of Eq.\,\eqref{eq:tcrit} we include two additional runs as well as a qualitative comparison with results from the works of \cite{Ayliffe2009II} and \cite{Szulagyi2016} in Appendix\,\ref{app:compare}. In all cases, we show that the criterion given by $t_{\rm crit}$ is satisfied. 

To illustrate the evolution of cooling time and rotational support, in Figure\,\ref{fig:rotational_support} (bottom row) we also show their simultaneous evolution in time and radius.
In the left panel, the color map shows the effective cooling time as a function of the planet's orbital time, $t$,  and radial distance from the planet, $r$. Initially, $t_{\rm c}$ rapidly increases to reach a monotonic growth phase. 
The right panel shows the same color map but for the specific angular momentum. After 30 orbits, the specific angular momentum continues to decrease for $r\lesssim 0.5 r_H$.
The decay of the specific angular momentum (even in nearly isothermal conditions) is due to the formation of an isentropic envelope at the smoothing length due to compressional heating from $P \nabla \cdot {\bf v}$ work. 
In this region, the gas pressure increases with time, with a more steep pressure gradient. 
As we will show in Section\,\ref{sec:radial_balance} this pressure gradient is enough to balance the planet's potential.

\subsubsection{Radial force balance}
\label{sec:radial_balance}
\begin{figure}
    \centering
    \includegraphics[scale=0.9]{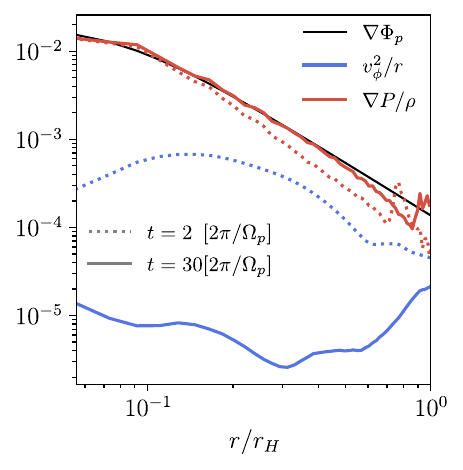}
    \caption{Spherical average of the radial pressure gradient (red) and $v^2_\phi/r$ (blue) for the \texttt{standard} run. The black solid curve shows $\partial_r \Phi_p$. Initially, pressure and rotational support are both important at scales $0.1r_H\lesssim r\lesssim r_H$. However, after local 30 orbits, the pressure support balances the gravitational acceleration at all scales inside the Hill sphere.}
    \label{fig:balance}
\end{figure}

\begin{figure}
    \centering
   \includegraphics[scale=0.9]{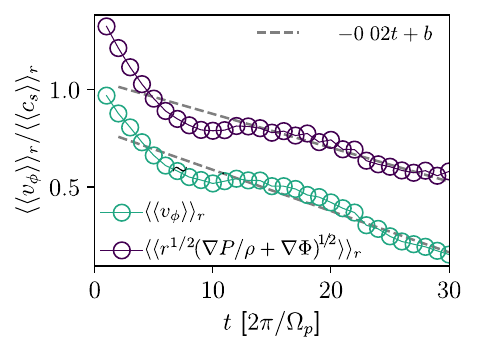}
    \caption{The plot shows the total average of the azimuthal velocity measured from the simulations (green circles) and estimated from radial force balance (purple circles) normalized by the sound speed. Both velocities decrease with time at the same rate. Results correspond to the the \texttt{standard} run. }
    \label{fig:balance_time}
\end{figure}

To better understand the coupled evolution of cooling and rotational support, we consider the radial force balance within the envelopes.
Initially,  all of the the simulations, regardless of thermodynamics,  show coherent rotational support at scales of $r\lesssim 0.3 r_H$. The average specific angular momentum of the in-falling material is usually some fraction of $r^2_H \Omega_p$ \citep{Ward2010}. As the sytem evolves, this specific angular momentum can be partially restored to the background protoplanetary disk by torques and/or advection of momentum. Moreover, the coupling between $v_\phi$ and $v_r$ through the radial momentum equation implies that some fraction of the angular momentum can also be delivered into radial momentum as the pressure gradient balances the gravitational force.
For example, consider a coordinate system centered at the planet's potential, the radial momentum equation corresponds to
\begin{equation}
    \partial_t (\rho v_r) + \nabla \cdot ( \rho v_r {\bf v} ) = \rho \left( \frac{v^2_\theta}{r} + \frac{v^2_\phi}{r}\right) - \partial_r P -\rho \partial_r \Phi_p \,.
    \label{eq:rhovr}
\end{equation}
where $\Phi_p = - GM_p/\sqrt{r^2 + R_s^2}$ is the smoothed potential of the planet.
Eq.\,\eqref{eq:rhovr} is coupled with continuity equations, energy flux, azimuthal and meridional momentum. 
Now, assume that initially, the system is nearly (but not perfectly) at rotational equilibrium, that is $v^2_\phi/r \sim \partial_r \Phi_p$. 
Since the gas inside the Hill sphere is not perfectly balanced, radial and meridional flows can deliver mass and entropy into the system, which in turn can increase the pressure support. 
As the pressure support increases to balance the gravitational acceleration, i.e., $\partial_r P / \rho \sim \partial_r \Phi_p$, the rotational support provided by $v^2_\phi/r$  will diminish, turning into radial momentum. Therefore, the system will evolve from a CPD-like morphology (partially rotationally supported gas flow) to a near spherically symmetric and pressure-supported envelope.  
To further test this hypothesis we consider the spherical average of the radial force equation. Neglecting viscous forces, the average radial force balance equation at the planet reference frame can be cast as
\begin{equation}
 0 \simeq \langle v^2_\theta/r \rangle + \langle v^2_\phi/r \rangle - \langle \partial_r P/\rho \rangle -  \partial_r \Phi_p \,,
 \label{eq:balancer}
\end{equation}
where $\Phi_p$ is the planet potential (including the smoothing length).
In Figure\,\ref{fig:balance} we show the the spherical average of the terms in Eq.\eqref{eq:balancer}. We omit the term $\langle v^2_\theta/r \rangle$ since is at least one order of magnitude smaller than $\langle v^2_\phi/r \rangle$. 
During the first planetary orbits,  $v^2_\phi/r$ provides additional support at scales $r\gtrsim 0.1r_H$. 
However, as time evolves, rotational support becomes negligible and the pressure gradient will balance the gravitational acceleration inside the Hill sphere. 
Moreover, the rate at which pressure support increases is the same rate as rotational support decreases. 
The rates are shown in Fig.\,\ref{fig:balance_time} where we plot the azimuthal velocity measured from the simulations together with the estimated velocity obtained assuming radial force balance, i.e., $v_{\phi} \simeq \sqrt{r (\partial_r P/\rho + \partial_r \Phi_p)} $. 
Both velocities decrease as a function of time and the obtained rate suggests that pressure balance should be fully reached after 50 planetary orbits. This rate indicates a short-lived rotationally supported envelope, however, this time scale is sensitive to the local physics of the disk. 

\subsection{Envelope structure in numerical models}

In this section, we describe the properties of the flow inside the Hill sphere for our simulations. 
We start by describing the velocity field for the \texttt{standard}, \texttt{multifluid} and \texttt{isothermal} models.
We also compare the results with a higher-resolution simulation to test the degree of numerical convergence.
We denote the higher-resolution model with \texttt{high\_res}. The model has the same initial conditions and parameters as the standard run, however, the mesh has been defined to achieve a resolution of $\sim 110\,\rm{cells}/r_H$, whereas the \texttt{standard} run has $\sim 55\,\rm{cells}/r_H$. 
Since we double the resolution, we also decrease the smoothing length for the \texttt{high\_res} model ($r_{\rm s} = 0.025 r_H$).

\subsubsection{Meridional velocity field and accretion}
\label{sec:meridional}

\begin{figure*}
    \centering
    \includegraphics[scale=0.9]{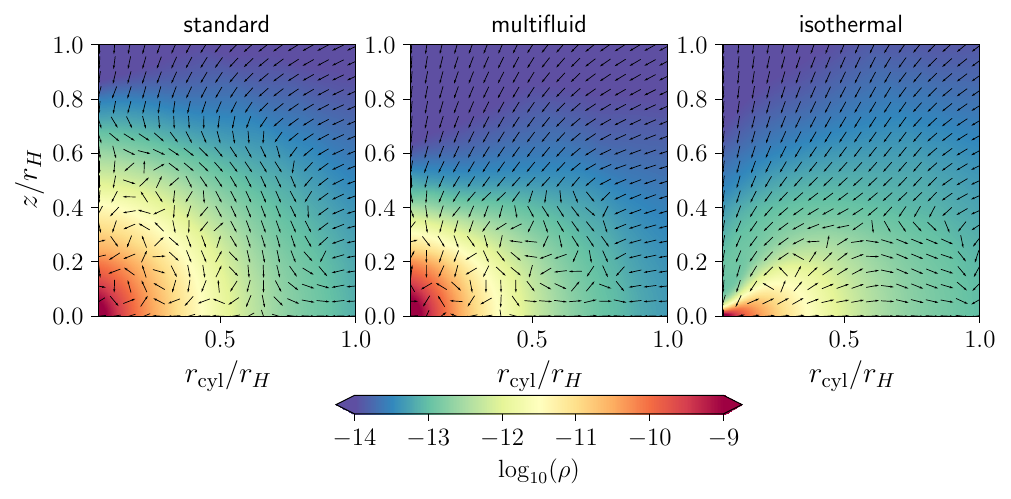}
    \caption{ Comparison between the \texttt{standard}, \texttt{multifluid} and \texttt{isothermal} models. The colorscale shows the density after 30 orbits, whereas the arrows indicate the azimuthal average of the meridional velocity field. The envelope in the \texttt{isothermal} run corresponds to a rotational-supported disk up to $r\lesssim r_H/3$, whereas the  \texttt{standard} and  \texttt{multifluid} runs show a spherically symmetric density distribution at scales $r\lesssim r_H/3$.   }
    \label{fig:fig_merid}
\end{figure*}
In Figure\,\ref{fig:fig_merid} we show the azimuthal average density inside the Hill sphere for the \texttt{standard}, \texttt{multifluid} and \texttt{isothermal} models at $t = 30 \, 2\pi/\Omega_p$. The results are displayed in cylindrical coordinates centered at the planet's potential. 
The \texttt{isothermal} run shows circumplanetary disks that extend out to $r \sim 0.3 r_H$. 
Contrary, the \texttt{standard} model displays a spherically symmetric density substructure that extends out to $r \sim 0.5 r_H$. 
The \texttt{multifluid} model, shows a similar spherically symmetric envelope as obtained in the \texttt{standard} model, however, the radial extension of the envelope in this case is about $r \sim 0.3 r_H$.

In Figure\,\ref{fig:fig_merid} we also include the azimuthal average of the meridional velocity field defined as $ {\bf v}_{\rm merid} = (v_{r_{\rm cyl}}, v_z) $, that is, the velocity field obtained from the average of the radial and vertical cylindrical velocities.
The \texttt{standard} model shows two distinctive patterns in the flow. At the inner regions $r\lesssim 0.5 r_H$, the mean flow is dominated by meridional circulation with a characteristic scale of $l \sim 0.1 r_H$.  

Beyond $r\sim 0.5 r_H$ the mean field has a vertical inflow that may penetrate the interior to $r\lesssim 0.3 r_H$ only near $z\simeq 0$. Away from that polar region, the flow circulates and leaves the Hill sphere at the midplane.

The meridional velocity field indicates that the envelope at $r\sim 0.5 r_H$ is partially shielded from the circulating material coming from the protoplanetary disks. 
This partially shielded interior region has been also reported in radiation hydrodynamical simulations \citep{Lambrechts2019} and adiabatic models with thermal relaxation \citep{Kurokawa2018}. 
We show in Section\,\ref{sec:envelope} that the gas inside this partially shielded region is an isentropic envelope in nearly hydrostatic equilibrium.
On the other hand, the material from the protoplanetary disks can reach the interior of the circumplanetary disks in the \texttt{isothermal} model. In this case, the meridional flow shows that polar inflow can reach the circumplanetary disks at scales $r\lesssim 0.1 r_H$, and material leaves the Hill sphere at the midplane \citep{Tanigawa2012,Kurokawa2018,Bethune2019,Fung2019}.

On average, the accretion towards the planet can be studied as a by-product of the meridional circulation. 
The mass flux through the Hill surface is obtained from the balance between mid-plane outflow and polar inflow, and is given by $\dot{m} = \langle r^2 \rho {\bf v}\cdot \hat{r} \rangle$, where $\hat{r}$ is the unit vector along $\bf{r}$. Inflow and outflow correspond to cases with ${\bf v} \cdot \hat{r} <0$ and ${\bf v} \cdot \hat{r} >0$, respectively.

In the \texttt{isothermal} model, we found that the outflow and inflow flux balances within 1\%-10\% at the scale of the Hill radius, in agreement with results reported in previous isothermal global models \citep{Fung2019,Krapp2022,Choksi2023}. 
Note, however, that the inclusion of a sink inside the smoothing length could lead to a different meridional flow pattern, but will not alter the accretion regime in a significant manner \citep{Choksi2023}. 
We discuss the need for more realistic boundary conditions in Section\,\ref{sec:discussion}.

At $r=r_H$ we report inflow and outflow values of $|\dot{m}|_{\rm in/out}\simeq 10\, M_J/{\rm Myr}$ with a net flux of $|\dot{m}|\simeq 0.5\, M_J/{\rm Myr}$ for the \texttt{isothermal} model.
Inside the Hill radius, the inflow/outflow rates increase giving $|\dot{m}|_{\rm in/out} \simeq 100\, M_J/{\rm Myr} $, with a net flux of $|\dot{m}| \simeq 10\, M_J/{\rm Myr}$ at $r=r_H$. 

Overall, the total mass increases during the first 5 orbits. After this short growing phase phase, the envelope has become rotationally supported. Subsequently, there is a density (mass) re-adjustment where the total mass of the envelope slowly decreases. 

In the \texttt{standard} and \texttt{multifluid} model, the mass flux is dominated by inflow at $r=r_H$ with a net flux of $\dot{m}\simeq 10 M_J/{\rm Myr}$. However, a transition to outflow-dominated flux occurs at $r=r_{\rm ise}$ in both cases, with a similar net flux $\dot{m} \simeq 10 M_J/{\rm Myr}$.
The total mass inside the Hill sphere increases as a function of time until pressure support fully balances the potential of the planet. The timescale for this mass adjustment is about $30$ and $60$ orbits for the \texttt{standard} and \texttt{multifluid} models, respectively. 
Overall, the flux of mass obtained in our simulations is in good agreement with previous work of \cite{Ayliffe2009I}, but somewhat lower than those reported in \cite{Szulagyi2014, Lambrechts2019}.
Note that our Jupiter mass planet is at a distance of  $30\,{\rm AU}$ while in previously mentioned works simulations adopt disk conditions at a distance of $\sim 5\,{\rm AU}$.
The choice of a different initial condition, disk viscosity, and numerical resolution complicates a detailed comparison between the measured accretion rates in physical units. We emphasize the good qualitative agreement of the meridional circulation patterns inside the Hill sphere.

\begin{figure*}[]
    \centering
    \includegraphics[scale=0.9]{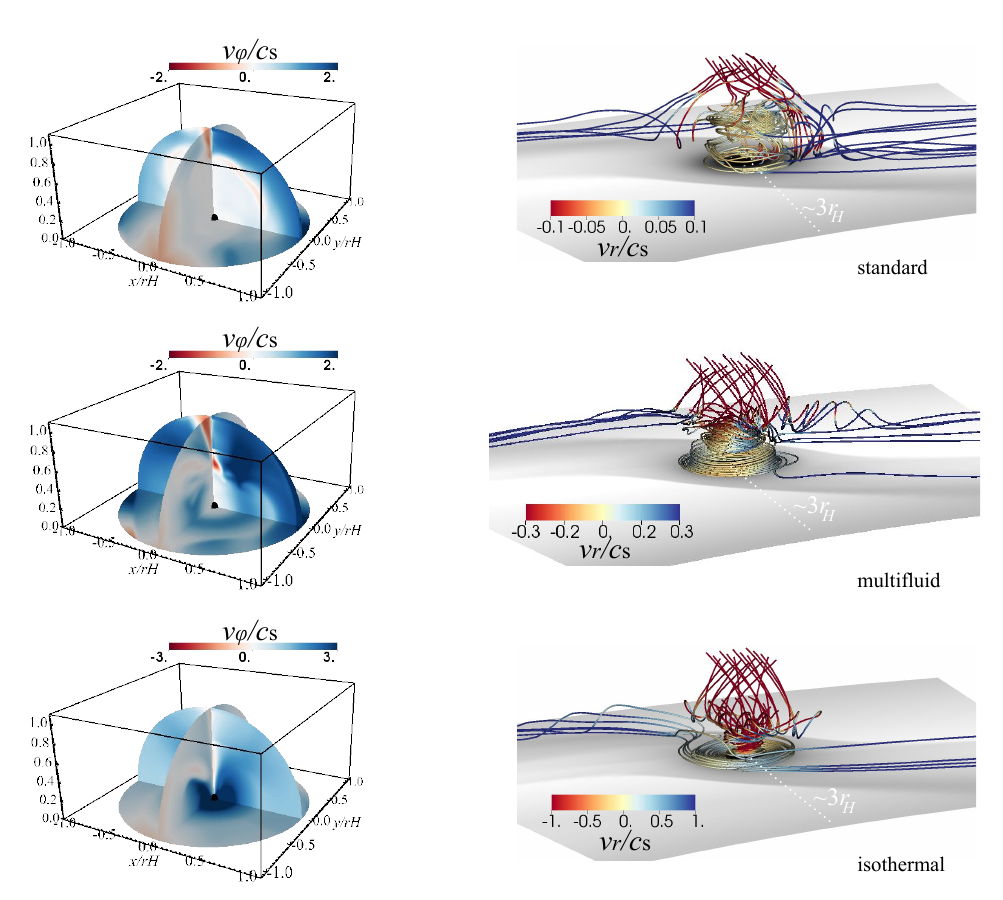}
    \caption{Comparison between the \texttt{standard} run (top panels), \texttt{multifluid} run (middle panels), and  \texttt{isothermal} model (bottom panels). The left panels show the azimuthal velocity normalized by the sound speed. The right panels show the streamlines of the velocity field at $30$ orbits. The flow has reached steady-state after 30 orbits. The color of the streamlines indicates accretion (red) and outflow (blue). We include the gas density at the midplane in gray color to highlight the location of the spiral wakes. In the isothermal case, the vertical inflow of material can reach the circumplanetary disks, however, the standard run shows circulation of the material above the midplane.    }
    \label{fig:fig_iso_stream}
\end{figure*}

\subsubsection{3D flow topology}
\label{sec:streams}

In Figure\,\ref{fig:fig_iso_stream} we compare the topology of the velocity field between the \texttt{standard}, \texttt{multifluid}, and \texttt{isothermal} models\footnote{The \texttt{low} model show similar results to the \texttt{multifluid} case, with a isentropic region inside $r\lesssim 0.1 r_H$ and weak rotational support between $0.1r_H \lesssim r \lesssim 0.3 r_H$.}.
The left panels of Figure\,\ref{fig:fig_iso_stream} show the azimuthal velocity inside the Hill sphere.
The azimuthal velocity in the \texttt{isothermal} highlights the presence of a CPD where  $r\lesssim0.3 r_H$. The supersonic azimuthal velocity indicates the strong rotational support at that scale. 
By contrast, the \texttt{standard} model shows negligible rotational support within the Hill Sphere. In turn, the flow is dominated by small-scale fluctuations, consistent with the small circulation pattern depicted in Figure\,\ref{fig:fig_merid} with the meridional velocity field. 

Intermediate between these regimes is the \texttt{multifluid} model, which still shows a coherent rotational velocity (net specific angular momentum) after $30$ orbits. While the gas is slowly evolving towards a spherically symmetric isentropic envelope as in the \texttt{standard} model, the transitory phase with faster cooling and higher net-specific angular momentum could promote the formation of a dusty sub-disk.  We defer a detailed discussion of the dust substructure to subsequent work (Paper II), and limit our discussion in this work to the critical cooling time and properties of the gas flow.  

The right panels of Figure\,\ref{fig:fig_iso_stream} show the streamlines of the velocity field after 30 planetary orbits in each model.
Within the Hill sphere, the systems have all reached a steady state in their flow field at this time, despite the aforementioned slow evolution in cooling time and specific angular momentum. 
In contrast, outside the Hill radius, the background protoplanetary disk continues to evolve as the planet carves a gap; however, gap opening seems to have little impact on the structure of the inner envelope. The simple expectation is a slow reduction of the mass flux towards the Hill surface \citep[][]{Lin1993}. 
In all models, the average inflow dominates near the polar regions, however, only in the \texttt{isothermal} model do the streamlines penetrate within $r\lesssim 0.3 r_H$.
The outflowing streamlines match up with the location of the spiral wakes excited by the planet. 

Of particular note is that in the models including radiative transfer, both inflow and outflow within the Hill Sphere are subsonic. This finding is in stark contrast to what is typically assumed in the core accretion model of giant planets in this mass range \citep{Lissauer:2011,Marleau2019,Marleau2023}. 
Supersonic flow is only seen in model \texttt{low}, where the opacity has been dramatically reduced to approach isothermal conditions. In this case, a weak shock forms above $r\sim 0.3 r_H$ as the flow reaches the CPD surface. 
Supersonic accretion at nearly the free-fall speed is only obtained for the local isothermal simulation. A dramatic reduction in the infall velocity could ultimately have implications for the assumed generation of accretion shocks, either onto CPDs or the planetary surface itself. We discuss in Section\,\ref{sec:envelope} other apparent discrepancies between our radiative transfer models and the standard assumptions of core accretion.

\subsubsection{Comparison with Spherically Symmetric Core Accretion Models: An isentropic structure?}
\label{sec:envelope}

To further compare our models with one-dimensional core accretion theory, we calculate the spherical average of the density, temperature, opacity, azimuthal velocity, entropy, and luminosity.
In Figure\,\ref{fig:envelope} we show the comparison between the runs \texttt{high\_res} and \texttt{standard} whereas in Figure\,\ref{fig:envelope_dust} we compare the runs \texttt{multifluid} and \texttt{standard}. 

We calculate the entropy: 
\begin{equation}
    S=c_v \ln( P/\rho_{\rm g}^\gamma ) / (k_B/m_H)\,,
\end{equation} 
where $k_B$ and $m_H$ are the Boltzmann constant and proton mass, respectively.

In all cases simulated in this work, the spherical average of the envelope is characterized by three different regions. 
An inner isentropic domain at $r\lesssim 0.2 r_H$, an outer isothermal domain at $r \geq 0.5 r_H$, and a transition between both isentropic and isothermal regimes where luminosity is nearly constant. 

As shown in Figure\,\ref{fig:fig_merid}, the isentropic region works as a barrier for the meridional flow \citep[see e.g.,][]{Kurokawa2018}. Whether this region is a numerical manifestation of the slowly contracting envelopes envisioned by \citet{Ginzburg2019} remains uncertain. Different boundary conditions at the scales of the smoothed potential could alter the formation of an isentropic inner envelope, though see \citep{Choksi2023}. 
One reason for the formation of the isentropic envelope is that the flow is not at equilibrium with the planet's potential initially. 
Therefore, the internal energy interior to the smoothing length increases as the material is compressed, adding heat to the system due to $P \nabla \cdot {\bf v}$ work.
Initially, this term introduces an energy rate between $\dot{E} \sim 10^{-8}-10^{-7} L_{\odot}$ at scales $r\lesssim r_{s}$ due to the converging flow towards the planet potential. After 20 orbits $\dot{E} \sim 10^{-5}-10^{-4} L_{\odot}$ as material continues to accrete near the smoothing length. This energy rate is sufficient to increase pressure support and form an isentropic region near the smoothed potential.
Note that in this work we do not include a source of energy inside the smoothing potential as accretion luminosity \citep{Benitez-Llambay2015}, which could further increase the cooling time inside the Hill sphere.

While the \texttt{multifluid} model shows an isentropic and nearly spherically symmetric envelope at scales of $r\lesssim 0.2r_H$, the streamlines and azimuthal velocity shown in Figure\,\ref{fig:fig_iso_stream} indicate the presence of rotation as well. We find that the dust within this region can successfully settle into an at least transitory CPD-like structure.
We will explore the properties of the dust component of the \texttt{multilfluid} model in Paper II. 
For the \texttt{low} run, the isentropic envelope forms at scales of $r\lesssim 0.1 r_H$. This result suggests that a more significant reduction of the opacity is required to rapidly cool the area near the smoothing length. 

\subsubsection{Hydrostatic Modeling}
After dividing the envelope into an isothermal and isentropic regime, the spherical average of the temperature and density can be modeled from solutions of the momentum equation in nearly hydrostatic equilibrium. As time evolves and the flow reaches a quasi-steady state the position of the isentropic and isothermal boundaries do not vary.
We denote the isentropic boundary as $r_{\rm ise}$.
In particular, inside the isentropic domain, the  temperature  can be described by the hydrostatic equilibrium solution 
\begin{equation}
    T(r) =  T_{\rm ise} +  \frac{G M_p}{\gamma c_v}  \left(\frac{1}{\sqrt{r^2+r_s^2}} -  \frac{1}{r_{\rm ise}}\right)\,,
\end{equation}
where the temperature at the isentropic radius, $r_{\rm ise}$, is $T_{\rm ise}$. 
The density is obtained from the isentropic condition $T\rho^{\gamma-1}=T_{\rm ise }\rho_{\rm ise}^{\gamma-1}$. 
The size of the isentropic region decreases as rotational support becomes important. 
It seems as well that the critical cooling time $t_{\rm c}  \gtrsim 0.1 \times 2\pi/\Omega_p$ is obtained at $r\simeq r_{\rm ise}$ for the \texttt{standard} and \texttt{multifluid} models.
This result implies that if one could estimate the value of $r_{\rm ise}$ from the initial disk conditions and the planet mass, one could also estimate the inner extension of the rotationally supported envelope. 
However, obtaining the value of $r_{\rm ise}$ from initial conditions will require a thorough numerical exploration with better-informed boundary conditions at the scales of the smoothed planet potential. 
Note that the outer extension of the circumplanetary disk is set by the tidal truncation radius $r_{\rm trunc}\sim r_H/3$ \citep{Martin2011}. In cases where $r_{\rm ise} \sim r_{\rm trunc}$ no rotational support is expected.
%

\subsubsection{Model Comparisons and Numerical Convergence}
\begin{figure*}
    \centering
    \includegraphics[scale=0.9]{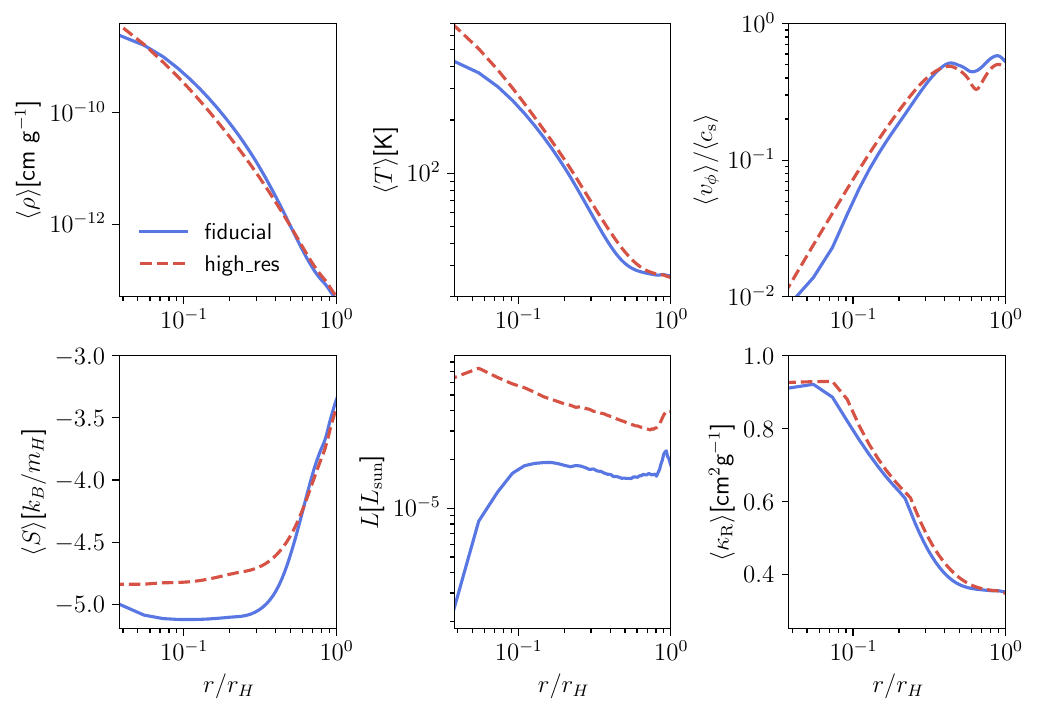}
    \caption{Spherical average of the density, temperature, azimuthal velocity, entropy, luminosity, and opacity inside the Hill hemisphere for the standard run (solid blue lines) at $t=20[2\pi/\Omega_p]$. The dashed red lines show the same average but for a run \texttt{high\_res} with $110\,\, {\rm cells}/r_H$, whereas the \texttt{standard} run has $55\,\,{\rm cells}/r_H$. }
    \label{fig:envelope}
\end{figure*}

Besides characterizing the spherical average of the planet's envelope, the results displayed in Figure\,\ref{fig:envelope} also show a high degree of convergence with numerical resolution. 
As expected, as we increase the resolution we reach hotter and denser regions of the envelope which can be seen at scales $r\lesssim0.1 r_H$. 
This hotter envelope shows a rise in the temperature of about $~20\%$ between $0.1r_H \lesssim r \lesssim 0.5 r_H$ in comparison to the \texttt{standard} model, as well as a slightly lower density at the same scales. 
These small difference in the density and temperature introduce only a small discrepancy of a factor $2-3$ between the luminosities of each model. The luminosities are shown in the bottom-middle panel of Figure\,\ref{fig:envelope}. 
The higher luminosity and small differences between temperature and density imply a slightly shorter cooling time for the higher-resolution run. However, the degree of rotational support is in very good agreement between the two models, as shown in the top-right panel of Figure\,\ref{fig:envelope}.

\begin{figure*}
    \centering
    \includegraphics[scale=0.85]{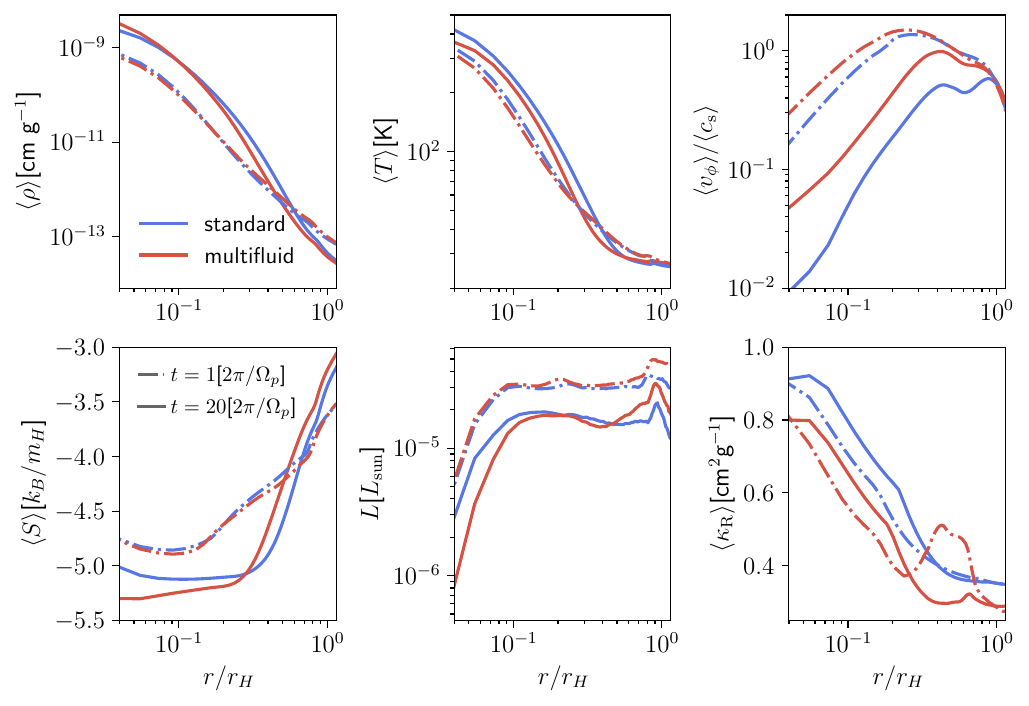}
    \includegraphics[scale=0.85]{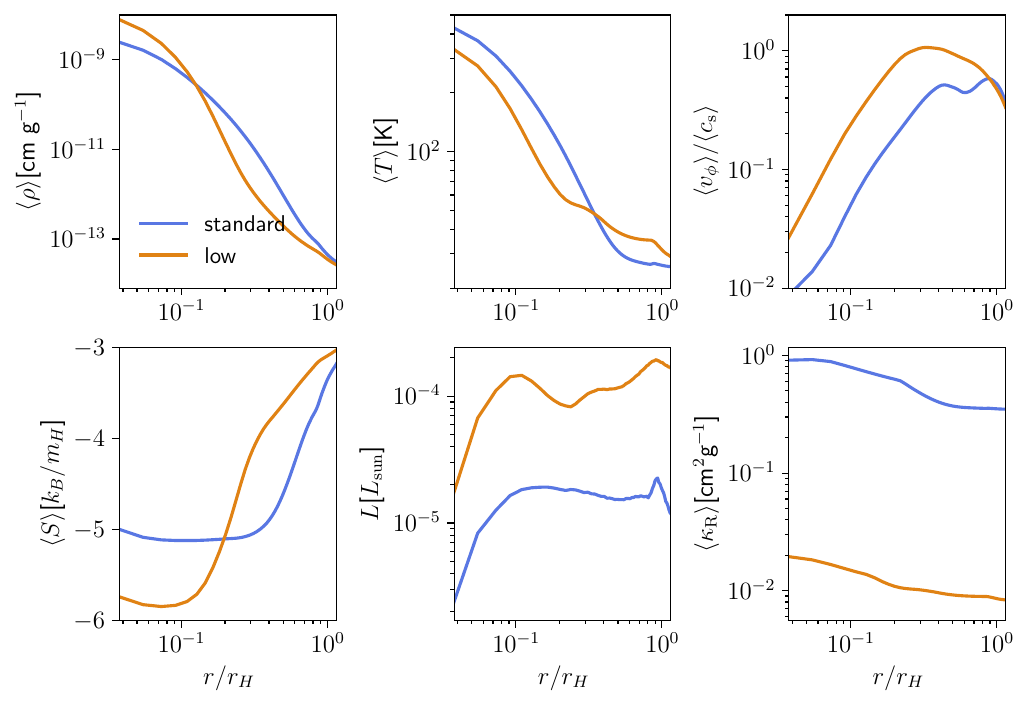}
    \caption{Spherical average of the density, temperature, azimuthal velocity, entropy, luminosity, and opacity inside the Hill hemisphere for the \texttt{standard} (solid blue lines), \texttt{mutlifluid} (solid red lines) and \texttt{low} (solid orange lines). The dotted line shows the results after 1 planet orbit, whereas solid lines correspond to 20 planet orbits. The \texttt{multifluid} model updates opacity with dust dynamics, whereas in the \texttt{low} model the opacity has been reduced $0.02$ of the \texttt{standard} value. }
    \label{fig:envelope_dust}
\end{figure*}

The same comparison is displayed in Figure\,\ref{fig:envelope_dust} for the \texttt{standard} and \texttt{multifluid} model. 
In this comparison, we include the results after 1 and 20 orbits, respectively. 
Both models show a similar time evolution, with density, and temperature increasing as the envelope reaches equilibrium. 
Both models obtain a similar average density, while the temperature and opacity are slightly smaller in the \texttt{multiluid} model.
Initially, the \texttt{multifluid} model also shows a slightly higher luminosity at scales $0.1 r_H \lesssim r \lesssim r_H$. 
These differences between average temperature, opacity, and luminosity translate into a shorter cooling time for the \texttt{multifluid} model. Simultaneously, rotational support is more prominent in the \texttt{multifluid} model. 
However, note as well that the ratio $\langle v_\phi \rangle/\langle c_s \rangle$ decreases as a function of time in both \texttt{standard} and \texttt{multifluid} model. 
In turn, both models show a nearly isentropic inner region.
While on average the Rosseland mean opacities are nearly the same between both models, the \texttt{multifluid} run shows an anisotropic distribution which introduces a (locally) shorter cooling time above the midplane (see Figure\ref{fig:fig1}). 
The spherical average shown in Figure\,\ref{fig:envelope_dust} does not highlight the faster cooling on the surface of the disk due to high opacity at the midplane.
Therefore, this anisotropic cooling favors rotational support. 
This result highlights the importance of self-consistent opacity modeling based on dust dynamics \citep{Krapp2022}. 
We have only considered here a single size distribution and we anticipate that reducing the mass load in the small grains will promote even faster cooling of the envelope. Therefore, the inclusion of dust evolution will also be  crucial to accurately evaluate the thermodynamic criterion for CPD formation.

\section{Discussion and Conclusions}
\label{sec:discussion}

The formation of circumplanetary disks has profound implications for the growth of the planet \citep{Pollack1996,Lubow1999}, the formation of natural satellites \citep{Canup2002,Mosqueira2003}, and the potential detection of giant protoplanets \citep{Isella2014,Benisty2021}.
Understanding the conditions for the formation of circumplanetary disks is therefore a crucial task for planet formation theories.
Thus, in this work, we provide a thermodynamic criterion for the formation of  rotationally supported envelopes around embedded planets.
We believe that our criterion based on effective cooling time (see Eq.\,\eqref{eq:tcrit}) synthesizes previous results from numerical simulations where circumplanetary disk formation is addressed in terms of opacity \citep{Ayliffe2009II,Schulik2020} and inner temperature cap \citep{Szulagyi2016,Szulagyi2017} (a comparison is shown in Appendix\,\ref{app:compare}.  
We expect that Eq.\,\eqref{eq:tcrit} is a necessary but not sufficient condition for the formation of circumplanetary disks, due to the formation of an inner isentropic envelope even with short cooling times.
Moreover, Eq.\,\eqref{eq:tcrit} can not be trivially cast in terms of disk initial conditions and planet mass. 
Therefore, to confirm the formation of circumplanetary disks we need to solve the radiation hydrodynamics equation to determine the evolution of specific angular momentum inside the Hill radius. 
While our simulations consider a particular disk model with a Jupiter mass planet, the thermodynamic criterion shown in Eq.\,\eqref{eq:tcrit} is agnostic to the planet mass and disk initial conditions. In other words, we anticipate that the correlation between cooling time and specific angular momentum displayed in  Figure\,\ref{fig:cooling_compare} is valid for a broad range of planetary mass, growing with different accretion rates and gap-opening time scales. This expectation is corroborated not only by the multipe runs in this work, but by comparisons with different numerical set-ups, planet, and disk parameters -- see the Appendix and Figure\,\ref{fig:comp_app}.

A common outcome of our non-isothermal models is the formation of an inner isentropic envelope, which is very weakly rotational. Typically, we find that the isentropic region extends out to $r \sim 0.2 r_H$, though the radial extension depends on the cooling conditions. This envelope structure has been observed in a variety of previous models in both the shearing sheet and global formalisms \citep{Szulagyi:2016,Lambrechts2017a,Zhu2021,Bailey2023}. Nevertheless, it remains unclear if this outcome captures an astrophysical phenomenon: heating from accretion onto a protoplanet, or if it is a consequence of choices of boundary conditions or thermodynamic approximations. We have verified that this isentropic region is not simply the result of insufficient resolution, which is consistent with the existence of the envelope in previous numerical work (e.g. \citep{Szulagyi:2016}) which achieved higher maximum resolution over short timescales.

In our simulations, we have identified that the main source of heat inside the smoothing length is due to the compressional heating term in the energy equation:  $P\nabla \cdot {\bf v}$. 
This energy input prevents rapid radiative cooling, and as time evolves it significantly contributes to the formation of the isentropic envelope, even in cases with reduced opacity of 2\% the standard value (\texttt{low} run). 
Because of the importance of compressional heating, the formation of the isentropic envelope is likely sensitive to the treatment of the inner boundary or planetary potential. The inflow rate of material from the disk to the Hill sphere, controlled by different disk models may also matter. For example, a different envelope could arise for the same planet-disk pair by adding a sink and/or considering a later stage of evolution with a deeper gap \citep{Choksi2023,Li2023}. Crucially, these different numerical or even physical choices should have a minor impact on the identified correlation between cooling time and specific angular momentum.

Besides heating from $P \nabla \cdot {\bf v}$, cells inside the smoothing length show modest heating from artificial viscosity since we do not use a tensor for the artificial viscosity as discussed in \citep{Stone1992}. Without a full treatment of the flow down to the planetary surface, including accurate thermodynamics and the self-gravity of the growing protoplanet envelope, it is not possible to determine whether this heating over or underestimates that which would be provided by a real growing planet.

Despite the uncertainties as to the origin of this isentropic region, it is consistent with some modern post-runaway models of slowly contracting giant planets \citep{Ginzburg2019}. Contrary to the rapid contraction model where the material is supplied by a circumplanetary disk \citep{Mordasini2012,Berardo2017}, the growth of a slow-contracting (nearly isentropic) envelope is driven by spherical accretion. 
Therefore, a better understanding of the formation of the isentropic region warrants further exploration, since we do not include additional heating sources such as accretion luminosity or shocks inside the smoothing length. 

Because of the lingering uncertainties in how to accurately correlate protoplanet energy input rates with astrophysical parameters (disk opacity, semi-major axis, protoplanet mass), the cooling criterion presented in this work is a useful benchmark for analytic and numerical models, that is agnostic to numerical choices. We stress that the formation of an inner isentropic envelope does not alter the correlation between specific angular momentum and cooling time described in this work. The correlation between specific angular momentum and cooling time is expected from an idealized model of a slow contracting unmagnetized gas cloud \citep{Nakamoto1994}. That is, a shorter cooling time could favor stronger rotational support that becomes dominant in the isothermal limit. 
In this work, we found the specific angular momentum is always smaller than that obtained in isothermal disk models. 
In our global simulations, envelopes only reach isothermal conditions at scales $r > r_H/3$, even in cases with reduced dust-to-gas mass ratio (see \texttt{low} model).
Thus, our results show that radiation hydrodynamics models with cooling times comparable to the orbital timescale bear little resemblance to isothermal models, even though this is a metric commonly invoked as ``nearly isothermal".

This cooling time scale, and therefore the rotational support of the gas envelope is sensitive to the opacity distribution. In this work, we have validated our thermodynamic criterion with multifluid simulations where the mean opacity is updated based on the local dust dynamics. 
We have considered standard-size distributions and settling-diffusion equilibrium. 
As shown in \cite{Krapp2022}, a different settling condition (and/or dust-to-gas mass load) impacts the opacity gradient.  
In cases where $100-500 \mu {\rm m}$ micron size grains are depleted,  due to e.g. dust evolution \citep{Birnstiel2012}, more anisotropic opacity distributions are expected at scales of $r \sim r_H/3$ and therefore shorter cooling times. In turn, rotational support could increase. 

Our global simulations with sufficient resolution at scales of the Hill radius also show that material is not in free-fall, but rather is characterized by  subsonic meridional circulation. Therefore, our global simulations can better inform the boundary conditions utilized in local models \citep{Berardo2017,Marleau2023}. Simultaneously, future global simulations could be improved by matching conditions of local models at the scale of the smoothing length (typically 100 to 1000 Jupiter radii). 
We stress, however, that a self-consistent model that connects the Hill sphere down to the core surface should also account for the gravitational potential and gravitational energy of the gas \citep{Bethune2019} as well as a more accurate equation of state \citep{DAngelo2013}. 
We have improved on existing models by including a  multifluid approach to solve the dust dynamics. 
Our multifluid model is crucial to accurately calculate the gas (material) opacity where temperatures are lower than $1500\,{\rm K}$, as well as, to characterize the distribution of solids (up to ${\rm cm}$-size particles) of the planet's envelope.

\section{Software and third party data repository citations} 

All the data was generated with a modified version of the open-source software FARGO3D available at https://bitbucket.org/fargo3d/public.git. The data underlying this article will be shared on reasonable request to the corresponding author.

\acknowledgments
We thank Yan-Fei Jiang for inspiring discussions and Gabriel-Dominique Marleau for useful comments and suggestions. We also thank the referee for a constructive, helpful report.
L.\ K.\   acknowledges support  from the Heising-Simons 51 Pegasi b postdoctoral fellowship,
L.\ K.\ ,  K.~M.~K and A.\ N.\ Y.\ acknowledge support  from  TCAN grant 80NSSC19K0639.
P.~B.~L. acknowledges  support from ANID, QUIMAL fund ASTRO21-0039 and FONDECYT project 1231205.
The results reported herein benefited from collaborations and/or information exchange within NASA’s Nexus for Exoplanet System Science (NExSS) research coordination network sponsored by NASA’s Science Mission Directorate and project “Alien Earths” funded under Agreement No. 80NSSC21K0593.
\appendix

\section{Additional runs}
\label{app:compare}
\begin{figure}
    \centering
    \includegraphics[scale=0.9]{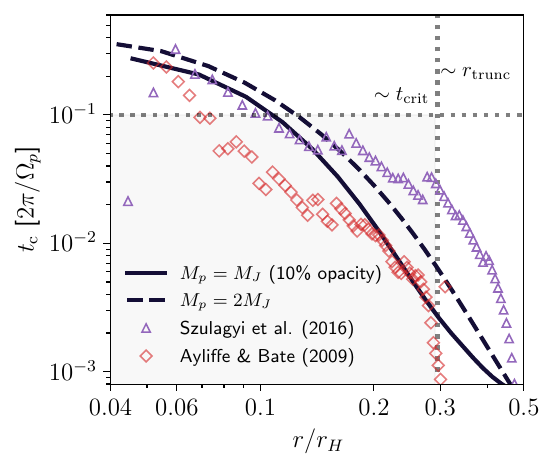}
    \caption{ Comparison of our cooling criterion with results from previous work of \cite{Ayliffe2009II} and \cite{Szulagyi2016} where the formation of circumplanetary disks has been reported. Cooling time is in good agreement with our criterion from Eq.,\eqref{eq:tcrit}.
    We have also extended our suite of runs to include a \texttt{standard} model but with a higher planet mass ($M_p = 2M_J$) and a reduced (0.1) opacity.  }
    \label{fig:comp_app}
\end{figure}

In this section, we validate the cooling criterion \eqref{eq:tcrit} including two additional runs with our setup from Section\,\ref{sec:numerical}. The first run considers a Jupiter mass planet and a reduced opacity with $\epsilon=0.0017$, whereas the second run corresponds to the \texttt{standard} model with a planet mass $M_p = 2 M_J$. 
In these two cases, we report cooling times shorter than $t_{\rm crit} = 0.1\, ( 2\pi/\Omega_p)$ and average specific angular momentum of $\langle j \rangle \sim 0.7 \langle j_{\rm iso} \rangle$ at scales of $r\sim 0.3 r_H$, providing additional support to our criterion. The cooling times obtained after 30 planet's orbits in our models are shown in Figure\,\ref{fig:comp_app}. In both cases, an isentropic envelope forms after 60 planetary orbits, and specific angular momentum decreases as reported for the \texttt{standard} run.

In Figure\,\ref{fig:comp_app} we additionally include the results from the previous works of \citep{Ayliffe2009II} and \citep{Szulagyi2016} and estimate the cooling time based on density and temperature profiles. 
In the case of \cite{Ayliffe2009II} we consider the case of $M_p = 100 M_{\rm earth}$ with a one-percent IGO reduced opacity 
(data obtained from dashed line profiles in Figure 2 from \cite{Ayliffe2009II}), whereas in the case of \cite{Szulagyi2016} we consider the case with $M_p=M_J$ and $T_p = 1500\, {\rm K}$ (blue curve in Figure 3 from \cite{Szulagyi2016}).
Both cases show the formation of circumplanetary disks, and therefore the effective cooling time should be shorter than $t_{\rm crit} = 0.1\, (2\pi/\Omega_p)$ if energy is predominantly transported by radiative diffusion.
There are several differences between our models and those from \cite{Ayliffe2009II} and \cite{Szulagyi2016}, therefore our comparison is meant to be qualitative. On the one hand, these previous works consider the planet at $5.2\,{\rm AU}$ from the central star where viscous heating is likely to set the temperature near the midplane. Moreover, \cite{Ayliffe2009II} also includes the self-gravity of the gas, which is neglected in our case. 
Note as well that the integration time scales, numerical resolution, and methods are also substantially different in all cases. 
Finally, since we do not have access to the opacity calculation, we estimate the opacity utilizing the table from \cite{Bell1994} (variations in the opacity may introduce discrepancies of a factor of a few in the cooling time). Overall, the cooling time is well below the $t_{\rm crit}$, which agrees with the formation of circumplanetary disks discussed in both works.

\bibliography{biblio,KMK_format_superbib}{}
\bibliographystyle{aasjournal}

\end{document}